\documentclass{conm-p-l}
\usepackage{amsfonts}
\usepackage{amsmath,amssymb}
\usepackage{latexsym}

\newcommand{\al}{\alpha}
\newcommand{\pa}{\partial}

\newcommand{\Si}{\Sigma}

\newcommand{\si}{\sigma}

\newcommand{\ta}{\tau}

\newcommand{\Om}{\Omega}

\newcommand{\De}{\Delta}

\newcommand{\rar}{\rightarrow}
\newcommand{\lrar}{\leftrightarrow}

\newcommand{\non}{\nonumber}

\copyrightinfo{2000}{American Mathematical Society}

\theoremstyle{definition}

\theoremstyle{remark}

\numberwithin{equation}{section}

\begin{document}

\title[Sutherland-type Trigonometric Models and Trigonometric Invariants]{Sutherland-type Trigonometric Models, Trigonometric Invariants and Multivariate Polynomials}

\author{K.G. Boreskov}
\address{Institute for Theoretical and Experimental Physics, Moscow 11259, Russia}
\curraddr{} \email{boreskov@itep.ru}
\thanks{Supported in part by grants RFBR 06-02-17012, 06-02-72041-MNTI and
SSh-843.2006.2 (Russia)}

\author{A.V. Turbiner}
\address{Instituto de Ciencias Nucleares, Universidad Nacional
Aut\'onoma de M\'exico, M\'exico DF 04510, Mexico}
\curraddr{Institut des Hautes Etudes Scientifique, Bures-sur-Yvette 91440, France}
\email{turbiner@nucleares.unam.mx, turbiner@ihes.fr}
\thanks{Supported in part by DGAPA grant IN121106 (Mexico)
and the University Program FENOMEC (UNAM, Mexico)}

\author{J.C. L\'opez Vieyra}
\address{Instituto de Ciencias Nucleares, Universidad Nacional
Aut\'onoma de M\'exico, M\'exico DF 04510, Mexico}
\curraddr{}
\email{vieyra@nucleares.unam.mx}
\thanks{Supported in part by DGAPA grant IN121106 (Mexico)}


\date{May 5, 2008}

\begin{abstract}
It is conjectured that any trigonometric Olshanetsky-Perelomov
Hamiltonian written in {\it Fundamental} Trigonometric
Invariants (FTI) as coordinates takes an algebraic form and preserves
an infinite flag of spaces of polynomials. It is shown that try-and-guess
variables which led to the algebraic form of trigonometric Olshanetsky-Perelomov Hamiltonians related to
root spaces of the classical $A_N, B_N, C_N, D_N, BC_N$ and
exceptional $G_2, F_4$ Lie algebras are FTI. This conjecture is also
confirmed for the trigonometric $E_6$ Olshanetsky-Perelomov
Hamiltonian whose algebraic form is found with the use of FTI.
\end{abstract}

\maketitle

\section{Introduction}

About 30 years ago, Olshanetsky and Perelomov \cite{Olshanetsky:1977} (for a review, see \cite{Olshanetsky:1983}) discovered a remarkable family of quantum mechanical Hamiltonians
with trigonometric potentials, which are associated to the crystallographic root spaces of the
classical ($A_N, B_N, C_N, D_N$) and exceptional ($G_2, F_4, E_{6,7,8}$) Lie algebras.
The Olshanetsky-Perelomov Hamiltonians have the property of complete integra-
bility (the number of integrals of motion in involution is equal to the dimension of
the configuration space) and that of exact solvability (the spectrum can be found
explicitly, in a closed analytic form that is a second-degree polynomial in the quantum numbers). The Hamiltonian associated to a Lie algebra $g$ of rank $N$, with root space $\Delta$, is
\begin{equation}
\label{H}
 { H}_\Delta = \frac{1}{2}\sum_{k=1}^{N}
 \left[-\frac{\pa^{2}}{\pa y_{k}^{2}}\right]\ +
 \frac{\beta^2}{8}\sum_{\alpha\in R_{+}}
 g^2_{|\alpha|}\frac{|\,\al|^{\,2}}
 {\sin^2 \frac{\beta}{2} (\alpha\cdot y)}\ ,
\end{equation}
where $R_+$ is the set of positive roots of $\Delta$,  $\beta\in \mathbb{R}$ is a parameter introduced for convenience,
$g^2_{|\alpha|}=\mu_{|\alpha|}(\mu_{|\alpha|}-1)$ are coupling
constants depending only on  the root length, and $y = (y_1, y_2,\ldots,y_N)$
is the coordinate vector. If all roots are of the same length, then
$g_{|\alpha|}=g$ (i.e. there is a single coupling constant). If the roots are of two
different lengths, then for the long roots $g_{|\alpha|}=g_l$ and for
the short ones $g_{|\alpha|}=g_s$ (i.e. there are two coupling constants).
The configuration space here is the Weyl alcove of the root space (see
\cite{Olshanetsky:1983}).

The ground state eigenfunction and its eigenvalue are
\begin{equation}
\label{Psi_0}
  \Psi_0 (y) \ =\ \prod_{\al\in R_+}
  \left|\sin \frac{\beta}{2} (\alpha\cdot y)\right|^{\mu_{|\al|}}\ ,\quad
  E_0\ =\ \frac{\beta^2}{8} \rho^2  \ ,
\end{equation}
where $\rho = \sum_{\alpha\in R_{+}} \mu_{|\alpha|} {\alpha}$ is the so-called `deformed Weyl vector' (see \cite{Olshanetsky:1983}, eqs.(5.5), (6.7)).
It is known that any
eigenfunction $\Psi$ has the form of (\ref{Psi_0}) multiplied by a polynomial
in   exponential (trigonometric) coordinates, i.e. $\Psi = \Phi \Psi_0$ (see
\cite{Olshanetsky:1983}). Such polynomials $\Phi$ are called
(generalized) {\it Jack polynomials}.
For connections between Jack polynomials, and the theory of special functions and orthogonal polynomials, see, e.g. \cite{Jack,Macdonald}.

\smallskip

For future use, we make three definitions.

\medskip

{\sc Definition 1.} A multivariate linear differential operator is said to be in algebraic form if its coefficients are polynomials in the independent variable(s). It is called algebraic if by an appropriate change of the independent variable(s), it can be written in an algebraic form.

\medskip

{\sc Definition 2}. Consider a finite-dimensional (linear) space of multivariate
polynomials defined as a linear span in the following way:
\[
 { P}^{(d)}_{n, \{\al \}} \ = \ \langle x_1^{p_1}
x_2^{p_2} \ldots x_d^{p_d} | 0 \leq \al_1 p_1 + \al_2 p_2 +\ldots +
\al_d p_d \leq n \rangle\ \ ,
\]
where the $\al$'s are positive integers  and $n\in \mathbb{N} $.
Its {\it characteristic vector} is the \hbox{$d$-dimensional} vector with components $\al_i$\footnote{We do not think that this notation will cause a confusion with positive roots.}:
\begin{equation}
 \vec \al = (\al_1, \al_2, \ldots \al_d)\ .
\end{equation}
For some characteristic vectors, the corresponding polynomial spaces may have a Lie-algebraic interpretation, in that they are the finite-dimensional
representation
spaces for some Lie algebra of (first-order) differential operators.
For example, the spaces corresponding to $\vec \al = (1, \ldots, 1)$, indexed by $n$, are finite-dimensional representation spaces of the algebra $gl(d+1)$ of  first-order differential operators.

\medskip

{\sc Definition 3.} Take the infinite set of spaces of multivariate polynomials
$P_n\equiv {P}^{(d)}_{n, \{\al \}}$, $n \in  \mathbb{N}$, defined as above, and order them by inclusion:
\[
{P }_0 \subset  { P}_1 \subset {P}_2 \subset \ldots
 \subset  {P}_n  \subset \ldots \ .
\]
Such an object is called an {\em infinite flag (or filtration)},
and is denoted ${P}^{(d)}_{\{\al \}}$. If a linear differential operator preserves
such an infinite flag, it is said to be  {\it exactly-solvable}. It is
evident that every such  operator is  algebraic (see \cite{Turbiner:1994}).
If the spaces $P_n$ can be viewed as the finite-dimensional representation spaces of some Lie algebra $g$, then $g$ is called the {\em hidden algebra} of
the exactly-solvable operator.

Any crystallographic root space $\De$ is characterized by its fundamental weights $w_a, a=1,2,\ldots r$, where $r={\rm{rank}}(\De)$. One can take a fundamental weight $w_a$
and generate its orbit $\Om_a$, by acting on it by all elements
of the Weyl group of $\Delta$. By averaging over this orbit, i.e. by computing
\begin{equation}
\label{Trig_Inv}
 \ta_{a}(y) = \sum_{\omega \in\Om_a} e^{i \beta (w \cdot y)}\ ,
\end{equation}
one obtains a trigonometric Weyl invariant for any specified $\beta\in \mathbb{R}$.
For a given root space $\Delta$ and a fixed $\beta$,
there thus exist $r$ independent trigonometric Weyl
invariants $\ta$ generated by $r$ fundamental weights $w_a$. We shall
call them {\it Fundamental} Trigonometric Invariants (FTI).
For the theory of root spaces, see \cite{Humphreys:1990} and
in a concise form, \cite{Ruehl:1999} or \cite{Sasaki:2000}.
A brief description of FTI, under the name
`exponential invariants' appears in Bourbaki \cite{Bourbaki:2002}, (Ch.6, $\S$3, p.194).

The goal of this paper is to show, for each of several  Lie algebras $g$, (i) that the Jack polynomials arising from the eigenfunctions of the Hamiltonian (\ref{H}), being rewritten in terms of FTI, remain polynomials in these invariants, (ii) that a similarity-transformed version of (\ref{H}), namely $h \propto \Psi_0^{-1} (H - E_0) \Psi_0$, acting on the space of trigonometric invariants (i.e., the space of trigonometric orbits) is an operator in algebraic form, and (iii) that $h$ preserves an infinite flag of spaces of polynomials, with a certain characteristic vector. Results are presented for the root spaces $A_N, BC_N, B_N, C_N, D_N, G_2, F_4$ and $E_6$. Although
similar results might seem to be obtainable for  $E_{7}$ and $E_{,8}$, an
analysis of those root spaces is absent, mainly due to great technical complications.

\medskip

\section{The case $\De = A_N$}

For the root space $A_N$, the Olshanetsky-Perelomov Hamiltonian
(\ref{H}) coincides with the Hamiltonian of the Sutherland model
\cite{Sutherland:1971}, and has the form
\begin{equation}
\label{H_AN}
    { H}_{\rm Suth}\ =\
  -\frac{1}{2}\sum_{k=1}^{N+1}\frac{\partial^{2}}{\partial x_{k}^{2}}
 + \frac{g \beta^2}{4}\sum_{k<l}^{N+1}\frac{1}{\sin^{2}(\frac{\beta}{2}(x_{k} -
 x_{l}))}\ ,
\end{equation}
with the ground state eigenfunction
\begin{equation}
\label{Psi_AN}
    \Psi_{0}(x)\ =\ \prod_{i<j}^{N+1}
 \sin^\nu\left(\frac{\beta}{2}(x_{i}-x_{j})\right)\ , \
 g=\nu(\nu-1)> - \frac{1}{4}\ .
\end{equation}
It describes a system of $(N+1)$ particles situated on a circle,
with a pairwise interaction that is given by potential term in
(\ref{H_AN}). For a review see \cite{Ruhl:1995}.

In order to solve the eigenvalue problem for the Hamiltonian
(\ref{H_AN}) let us introduce the Perelomov relative coordinates
\cite{Perelomov:1971}
\begin{equation}
\label{Pere}
Y\ =\ \sum x_i\ ,\ y_i\ =\ x_i - \frac{1}{N+1} Y\ ,\
i=1,\ldots , N+1  \ ,
\end{equation}
where $Y$ is the center-of-mass coordinate, and the coordinates $y_i$
are confined to the hyperplane
\begin{equation}
   \sum_{i=1}^{N+1} y_i\ =\ 0 \ .
\end{equation}
A transformation to Weyl-invariant periodic coordinates was introduced in
\cite{Ruhl:1995}. It is
\begin{equation}
\label{RT_A}
    (x_1,x_2,\ldots x_{N+1}) \rightarrow \big(e^{i\beta Y},\eta_n(x)=\sigma_n(e^{i
\beta y(x)})| \ {n= 1,2 \ldots N}\big)  \ ,
\end{equation}
where $\sigma_{k}(x) = \sum_{i_{1}<i_{2}<\cdots<i_{k}}
x_{i_{1}}x_{i_{2}}\cdots x_{i_{k}}, \ k=1,2,\ldots, N$ are the
elementary symmetric polynomials, and by convention, $\si_0=\si_{N+1}=1$
and $\si_i=0$ for $i<0$ and $i>(N+1)$.
It was shown that a similarity-transformed version of the Hamiltonian (\ref{H_AN}),
namely $h_{A_N} = -\frac{2}{\beta^2}(\Psi_{0})^{-1}\,
{H_{Suth}}\,\Psi_{0}$\,, after separation of the center-of-mass motion
$(Y=0)$ takes on the algebraic form
\begin{equation}
\label{h_AN}
 {h}_{\rm Suth} = \sum_{i,j=1}^{N}{ A}_{ij}(\eta)
\frac{\partial^2}{\partial {\eta_i} \pa {\eta_j} } +
 \sum_{i=1}^{N}{ B}_i(\eta)
\frac{\partial}{\partial \eta_i} \ ,
\end{equation}
where
\[
{ A}_{ij}\ =\ \frac{(N+1-i)\,j}{N+1}\,\eta_{i}\,\eta_j +
 \sum_{{l\geq}{\max (1,j-i)}} (j-i-2l)\,\eta_{i+l}\,\eta_{j-l}\quad
 \  \mbox{at}\ i\geq j\ ,
\]
\[
 \
  {A}_{ji}\ = {A}_{ij}\ , \  {B}_i\  =\ (\frac{1}{N+1}+\nu)\,i\,(N+1-i)\,\eta_{i}\ .
\]
It can  easily be checked \cite{Ruhl:1995} that the operator ${h}_{\rm Suth}$
preserves the infinite flag ${P}^{(N)}_{\{ 1,1,\ldots, 1 \}}$. This is in
agreement with our general conjecture that the characteristic vector
for the trigonometric model coincides with the minimal
characteristic vector for corresponding rational model
\cite{BLT-rational}. The operator $h_{A_N}$  depends on a
single parameter $\nu$, linearly. The nodal structure of its eigenpolynomials
(i.e. where they vanish) at fixed $\nu$ remains an open question.

\medskip

{\sc Statement 1.} {\it For any $n$, one can find a fundamental
weight $w_a$ of the $A_N$ root system for which  $\eta_{n}=\tau_a$. Hence,
the Weyl-invariant periodic coordinates $\eta_{n}, n=1,2,\ldots,N$, defined in
(\ref{RT_A}), coincide with the fundamental trigonometric invariants
$\tau_a, a=1,\ldots, N=$rank~(${A_N}$), defined in  (\ref{Trig_Inv}).}

\medskip

To prove this statement,  note that the fundamental weights of $A_N$ can be written in terms of the canonical basis $e_1,\ldots, e_{N+1}$ of $\mathbb{R}^{N+1}$ as (see \cite{Bourbaki:2002})
\begin{equation}
 w_k = (e_1 +e_2+\dots +e_k) -\frac{k}{N+1}\sum_{j=1}^{N+1}e_j ~,
 \quad k=1,N \ .
\end{equation}
Hence, the orbit element related to a given fundamental weight reads at $Y=0$ as
\begin{align}
\label{S1}
 \exp(i\beta w_k\cdot y) = \exp\left(i\beta \sum_{j=1}^k
  y_j\right)=\prod_{j=1}^k \exp(i\beta y_j) \ .
\end{align}
Since the Weyl group for $A_N$ is a symmetric group $S_{N+1}$ that permutes the vectors $e_j$, the averaging of \eqref{S1} over this group gives for the FTI exactly $\sigma_k(\exp(i\beta y))$. It is worth noting that there exists a
symmetry \cite{Ruhl:1995}: the involution $\beta \lrar -\beta$ corresponds to
$\eta_i \lrar \eta_{N+1-i}$ (see \eqref{S1}). Since the original Hamiltonian depends on $\beta^2$, this leads to  certain relations between the coefficients
$A_{ij}$ and $B_i \lrar B_{N+1-i}$.

Let us consider the algebra $gl(N+1)$  realized by the first order
differential operators

\begin{subequations}
\label{Jops}
\begin{eqnarray}
 { J}_i^- &=& \frac{\pa}{\pa \tau_i},\qquad \quad
i=1,2\ldots N \ ,  \\
 {{ J}_{ij}}^0 &=&
\tau_i \frac{\pa}{\pa \tau_j}, \qquad i,j=1,2\ldots N \ ,
  \\
{ J}^0 &=& \sum_{i=1}^{d} \tau_i\frac{\pa}{\pa \tau_i}-n\, ,
  \\
 { J}_i^+ &=& \tau_i { J}^0 =
\tau_i\, \left( \sum_{j=1}^{d} \tau_j\frac{\pa}{\pa \tau_j}-n
\right), \quad i=1,2\ldots N \ ,
\end{eqnarray}
\end{subequations}
where $n$ is any number. If in (\ref{Jops}), $n$ is non-negative
integer, the generators (\ref{Jops}) will have a common invariant subspace
${P}^{(N)}_{n,\{ 1,1,\ldots, 1 \}}$, on which they act irreducibly.
Hence the infinite flag ${P}^{(N)}_{\{ 1,1,\ldots, 1 \}}$ is made of
irreducible finite-dimensional representation spaces of the algebra
$gl_{N+1}$. If the raising generators $J^+$  are excluded, the
remaining generators will form the maximal affine subalgebra of the
$gl(N+1)$ algebra. It is evident that the generators $J^0, J^-$ preserve
 ${P}^{(N)}_{\{ 1,1,\ldots, 1
\}}$. It can be proved that ${h}_{\rm Suth}$, given in (\ref{h_AN}) can be
rewritten in terms of the generators $J^0, J^-$.
Therefore, $gl_{N+1}$ is the hidden algebra of the Sutherland model.

\medskip

\section{The case $\De = BC_N$ (including $\De = B_N, C_N, D_N$)}

For the root space $BC_N$ the Olshanetsky-Perelomov Hamiltonian
(\ref{H}) has the form
\begin{eqnarray}
\label{H_BCN}
 { H}_{BC_N} &=&
-\frac{1}{2}\sum_{i=1}^{N} \frac{\pa^2}{\pa {x_i}^2} \!+
\frac{g\beta^2}{4} \sum_{i<j}^{N} \left[
\frac{1}{\sin^2\!\left(\frac{\beta}{2}(x_{i}-x_{j})\right)} +
\frac{1}{\sin^2\!\left(\frac{\beta}{2}(x_{i}+x_{j})\right)} \right]
\nonumber \\ && + \ \frac{g_{2}\beta^2}{2} \sum_{i=1}^{N}
\frac{1}{\sin^2\! \beta x_{i} } + \ \frac{g_{3}\beta^2}{8}
\sum_{i=1}^{N} \frac{1}{\sin^2\! { \frac{\beta x_{i}}{2}}} \ .
\end{eqnarray}
with the ground state eigenfunction
\begin{equation}
\label{Psi_BCN}
 \Psi_{0} \ =\ \prod_{i<j}^N |\sin(\frac{\beta}{2} (x_{i}-x_{j}))|^{\nu}
 |\sin(\frac{\beta}{2} (x_{i}+x_{j}))|^{\nu}
 \prod_{i=1}^N |\sin(\beta x_i)|^{\nu_{2}} |\sin({\frac{\beta}{2}x_i})|^{\nu_3} \ ,
\end{equation}
where $g = \nu(\nu - 1)>-1/4\ ,\ g_{2} = \nu_{2}(\nu_{2} - 1)>-1/4\, , \ g_{3}
= \nu_{3}(\nu_{3} + 2\nu_{2} - 1)>-1/4\,$. From the general $BC_{N}$
Hamiltonian (\ref{H_BCN}) the $B_{N}$, $C_{N}$ and $D_{N}$ cases are
obtained by specializing as follows:
\begin{itemize}
\item $B_{N}$ case:  $\nu_{2}=0$,
\item $C_{N}$ case:  $\nu_{3}=0$,
\item $D_{N}$ case:  $\nu_{2}=\nu_{3}=0$.
\end{itemize}

In order to solve the eigenvalue problem for the $BC_{N}$
Hamiltonian (\ref{H_BCN}), let us perform a change of variables to Weyl-invariant periodic coordinates \cite{Brink:1997}, i.e.,
\begin{equation}
\label{RT_BC}
    (x_1,x_2,\ldots x_{N}) \rightarrow \big(\eta_n(x)=\sigma_n(\cos{\beta x})|
    \ {\scriptstyle n= 1,2 \ldots N}\big)
\end{equation}
where $\sigma_{k}(x)\ , k=1,2,\ldots, N$ are the elementary
symmetric polynomials, with \hbox{$\si_0=1$}. The
similarity-transformed Hamiltonian (\ref{H_BCN}): $$h_{BC_N} =
-\frac{2}{\beta^2}(\Psi_{0})^{-1}\, {H_{BC_N}}\, \Psi_{0}$$ with
$\Psi_{0}$ given by (\ref{Psi_BCN}), has the  form
(see \cite{Brink:1997})
\begin{equation}
\label{h_BCN}
 {h}_{BC_N}\ =\ \sum_{i,j=1}^{N} {A}_{ij}({\eta})
\frac{\pa^2}{\pa {{\eta}_i} \pa {{\eta}_j} } +
\sum_{i=1}^{N}{B}_i({\eta}) \frac{\pa}{\pa {\eta}_i} \ ,
\end{equation}
where
\begin{eqnarray}
 {A}_{ij} &=&
N\,{\eta}_{i-1}\,{\eta}_{j-1} - \, \sum_{l\ge 0} \Big[
  (i-l)   \,{\eta}_{i-l}  \,{\eta}_{j+l}
+ (l+j-1) \,{\eta}_{i-l-1}\,{\eta}_{j+l-1} \non
\\ & &
-(i-2-l) \,{\eta}_{i-2-l}\,{\eta}_{j+l} - (l+j+1)
\,{\eta}_{i-l-1}\,{\eta}_{j+l+1} \Big] \ , \non\\
 {B}_i  &=&
\nu_3(i-N-1)\,{\eta}_{i-1} - \Big[1 + \nu(2N-i-1) + 2 \nu_2 +
\nu_3\Big]\,i\,{\eta}_{i} \non
\\ & &
 - \nu(N-i+1)(N-i+2){\eta}_{i-2} \ . \non
\end{eqnarray}
Here $\eta_0=1$ and by convention, $\eta_i=0$ for $i<0$ and $i>N$.

It can be easily checked that the operator ${h}_{BC_N}$ preserves
the infinite flag ${P}^{(N)}_{\{ 1,1,\ldots, 1 \}}$, similarly to the case
 of the $A_N$ Hamiltonian. This is in agreement with our
conjecture that the characteristic vector for a trigonometric model always coincides
with the minimal characteristic vector for the corresponding rational model~\cite{BLT-rational}.

The $BC_N$ model depends on three parameters $\nu, \nu_2, \nu_3$, and
the nodal structure of the eigenpolynomials (i.e. where they vanish) at fixed
$\nu$'s remains an open question.

\smallskip

{\sc Statement 2:} {\it For any $n$, one can find a fundamental
weight $w_a$ of the $C_N$ root system for which $\eta_{n}= f_a \tau_a$,
where $f_a$ is a constant. Hence, the
Weyl-invariant periodic coordinates $\eta_{n}, n=1,\ldots,N$
(\ref{RT_BC}) coincide with the fundamental trigonometric invariants
$\tau_a, a=1,\ldots,$ \em N=rank~(${C_N}$), defined in  \it (\ref{Trig_Inv})}, up
to numerical factors. The coefficients in (\ref{h_BCN}) are changed accordingly.

Indeed, the element of the $k$-th orbit related to the fundamental weight
$ w_k = (e_1 +e_2+\dots +e_k)\ , \quad k=1,2,\ldots N)$ (see \cite{Bourbaki:2002}), looks like
\begin{align}
\label{S2}
\exp(i\beta w_k\cdot x) = \prod_{j=1}^k \exp(i\beta x_j) ~.
\end{align}
The Weyl group for $C_N$ root space is a semidirect product of a permutation group $S_N$ acting on the vectors $e_i$ and a group $(\mathbb{Z}/2\mathbb{Z})^N$ that acts as $e_j\mapsto (\pm 1)_j e_j$.
Averaging \eqref{S2} over the second group action gives $2^k\prod_{j=1}^{k}\cos(\beta x_j)$, and averaging over permutations gives $\sigma_k(\cos(\beta x))$, up to a common
multiplicative factor.

\smallskip

{\sc Remark 1:} The set of  $C_N$ trigonometric invariants of the form
(\ref{Trig_Inv}) is characterized by the smallest common period, in
comparison with the set of the $B_N$ or $D_N$ trigonometric invariants.
In general, any $C_N$ trigonometric invariant can be rewritten as a polynomial either in $B_N$ or in $D_N$ invariants.

\smallskip

{\sc Remark 2:} Neither the $B_N$ Hamiltonian ($g_{2}=0$ in
(\ref{H_BCN})) nor the $D_N$ Hamiltonian ($g_{2}=g_3=0$ in
(\ref{H_BCN})) takes on an algebraic form in terms of the $B_N$ or
$D_N$ trigonometric invariants, respectively. However, both $B_N$
and $D_N$ Hamiltonians take on an algebraic form in terms of the $C_N$
trigonometric invariants.

\smallskip

It can be shown that the operator ${h}_{\rm BC_N}$ of (\ref{h_BCN})
can be rewritten in terms of the generators $J^0, J^-$ of (\ref{Jops})
(see \cite{Brink:1997}), similarly to the $A_N$ model; cf. (\ref{h_AN}). Therefore,   $gl_{N+1}$ is the hidden algebra
of the $BC_N$ model.

\section{The case $\De = G_2$}

The Olshanetsky-Perelomov Hamiltonian (\ref{H}) for the root space
$G_2$ has the form
\[
{ H}_{\rm G_2} =
 -\frac{1}{2}\sum_{k=1}^{3}\frac{\pa^{2}}{\pa x_{k}^{2}}
 + \frac{g \beta^2}{4}\sum_{k<l}^{3}\frac{1}{\sin^{2}
 (\frac{\beta}{2}(x_{k} - x_{l}))}
\]
\begin{equation}
\label{H_G2}
 + \frac{g_1 \beta^2}{4}\sum_{ k<l ; k,l \neq m}^{3}
 \frac{1}{\sin^{2}(\frac{\beta}{2}(x_{k} + x_{l}-2x_{m}))}\ ,
\end{equation}
where $g=\nu (\nu-1) > -\frac{1}{4}$ and $g_1=3\mu (\mu -1) >
-\frac{3}{4}$ are the coupling constants associated with
two-body and three-body interactions, respectively. From a
physical point of view, the Hamiltonian (\ref{H_G2}) describes a
system of three identical particles that are situated on a circle.
The ground state eigenfunction is
\begin{equation}
\label{Psi_G2}
 \Psi_{0}(x) = (\De^{(trig)}(x))^{\nu}
(\De_1^{(trig)}(x))^{\mu}\ ,
\end{equation}
where $\De^{(trig)}(x),\ \De_1^{(trig)}(x)$ are the trigonometric
analogies of the Vandermonde determinant and are defined by
\begin{subequations}
\begin{eqnarray}
\De^{(trig)} (x) &=& \prod_{k<l}^3 |\sin\frac{\beta}{2}(x_{k}-x_{l})| \, ,\\
\De_1^{(trig)} (x) &=& \prod_{ k<l ; k,l \neq m}^3|
\sin\frac{\beta}{2}(x_{k}+x_{l}-2x_{m})| \ .
\end{eqnarray}
\end{subequations}
In order to solve the eigenvalue problem for the Hamiltonian
(\ref{H_G2}), let us introduce the Perelomov coordinates $Y,
y_i,\ i=1,2,3$ as in (\ref{Pere}). The relative coordinates $y_i$ are
constrained by \(y_1 + y_2 + y_3\ =\ 0 \ \).
It was shown in \cite{Rosenbaum:1998} that after separating the
center-of-mass coordinate in (\ref{H_G2}), and introducing the Weyl-invariant
periodic coordinates
\begin{subequations}
\begin{eqnarray}
\eta_1  &=&
\frac{-2}{\beta^2}\bigg[\sin^2\frac{\beta}{2}(y_1-y_2)+\sin^2\frac{\beta}{2}(
y_2-y_3)+\sin^2\frac{\beta}{2}(y_3-y_1) \bigg] ,
\label{RT_G} \\
 \eta_2  &=&  \frac{4}{\beta^6} \biggl[ \sin\beta(y_1-y_2)+\sin\beta(
y_2-y_3)+\sin\beta(y_3-y_1)\biggr]^2 \ ,
 \end{eqnarray}
\end{subequations}
the similarity-transformed Hamiltonian $h_{{G}_2} =
-2(\Psi_{0})^{-1}\, { H}_{G_2}\, \Psi_{0}$ takes, after the
transformation to new coordinates
$(x_1,x_2,x_3) \rightarrow \big(Y,\ \eta_1,\ \eta_2\big)$ the
algebraic form
\[
h_{G_2} (\eta)\ =\
 -\biggl(2\eta_1+\frac{\beta^2}{2}\eta_1^2-\frac{\beta^4}{24}\eta_2\biggr)
 \pa_{\eta_1\eta_1}^2 - \biggl(12+\frac{8\beta^2}{3}\eta_1\biggr)\eta_2
 \pa_{\eta_1\eta_2}^2
\]

\[
 + \biggl(\frac{8}{3}\eta_1^2\eta_2-2\beta^2\eta_2^2\biggr)
 \pa_{\eta_2\eta_2}^2
      - \biggl\{2[1+3(\mu +\nu)]+\frac{2}{3}(1 + 3\mu +
       4 \nu)\beta^2\eta_1\biggr\}\pa_{\eta_1}
\]

\begin{equation}
\label{h_G2_eta}
 +\bigg\{\frac{4}{3}(1+2\nu)\eta_1^2-[\frac{7}{3}
        +4(\mu +\nu )]\beta^2\eta_2\bigg\}
                \pa_{\eta_2}\ .
\end{equation}
It can  easily be checked that the operator ${h}_{G_2}$ preserves the
infinite flag ${P}^{(2)}_{\{ 1,2 \}}$. This is in agreement with our general
conjecture that the characteristic vector for a trigonometric model coincides
with the minimal characteristic vector for the corresponding rational model
\cite{BLT-rational}.

The root space $G_2$ has two fundamental weights; namely, $a_1=e_3-e_1$ and
$a_2=-e_1-e_2+2e_3$. Averaging over the orbits generated by $a_1$
and $a_2$ (as in (\ref{Trig_Inv})), we end up with the explicit FTI
\begin{subequations}
 \begin{eqnarray}
\label{tau1_G2}
 \tau_1\ &=&\ 2[\cos(\beta(y_1-y_2)) +
 \cos(\beta(2y_1+y_2)) + \cos(\beta(2y_1+y_2))]\ , \\[5pt]
\label{tau2_G2}
 \tau_2 &=& 2[\cos(3\beta y_1) +
 \cos(3\beta y_2) + \cos(3\beta(y_1+y_2))]\ .
\end{eqnarray}
\end{subequations}
One can easily verify a connection between
$\eta_{1,2}$ and $\tau_{1,2}$:
\begin{equation}
\label{transf}
  \eta_1 = \frac{1}{2\beta^2} (\tau_1-6)\ ,\
  \eta_2 = \frac{1}{\beta^6} (4\tau_2-\tau_1^2+12)\ .
\end{equation}
The transformation (\ref{transf}) does not alter the infinite flag
${P}^{(2)}_{\{ 1,2 \}}$. Changing the variables in (\ref{h_G2_eta})
from $\eta$'s to $\tau$'s we end up with
\[
h_{G_2} (\tau)\ =\ \frac{1}{8\beta^2} h_{G_2}\ =\
\]
\[
 \biggl(4 + \tau_1+\frac{\tau_2}{3}-\frac{\tau_1^2}{3}\biggr)
 \pa_{\tau_1\tau_1}^2
 -\biggl(12+4\tau_2+\tau_1\tau_2-2\tau_1^2\biggr)
 \pa_{\tau_1\tau_2}^2
 -\biggl(9\tau_1 + 3\tau_2 + 3 \tau_1\tau_2 + \tau_2^2 - \tau_1^3 \biggr)
 \pa_{\tau_2\tau_2}^2
\]
\begin{equation}
\label{h_tau_G2}
  +\biggl[2\nu -\frac{1 + 3\mu + 4 \nu}{3}\tau_1\biggr]\pa_{\tau_1}
 -\bigg[ 3(2\mu+\nu)+(1+2\mu+2\nu)\tau_2+\frac{\nu}{12}\tau_1^2 \bigg]
                \pa_{\tau_2}\ .
\end{equation}
A straightforward analysis confirms a conclusion that the operator ${h}_{G_2}
(\tau)$ preserves the infinite flag ${P}^{(2)}_{\{ 1,2 \}}$. This is not a surprising
result, since the transformation (\ref{transf}) maps each subspace
in  ${P}^{(2)}_{\{ 1,2 \}}$ to itself.

The $G_2$ model depends on two parameters $\nu, \mu$, and
the nodal structure of eigenpolynomials (i.e. where they vanish) at fixed
$\nu$ and $\mu$ remains an open question.

\medskip

Let us consider an infinite-dimensional Lie algebra of the differential
operators generated by the following eight operators
\begin{equation}
\label{g2_gen}
\left\{
\begin{array}{ll}
L^1  =  \pa_{\tau_1}\ ,                    &   L^2  = \tau_1\pa_{\tau_1} - {n\over 3}\ ,
\\[5pt]
L^3  =  2\tau_2\pa_{\tau_2}- {n\over 3}\ , &  L^4 =\tau_1^2\pa_{\tau_1}+2\tau_1\tau_2
\pa_{\tau_2} - n\tau_1 \ ,  \\[5pt]
L^5  = \pa_{\tau_2}\ ,          &  L^6  = \tau_1\pa_{\tau_2} \ , \\[5pt]
L^7  =  \tau_1^2\pa_{\tau_2}\ , &  T    = \tau_2\pa_{\tau_1\tau_1}^2 \ .
\end{array}
\right.
\end{equation}

This algebra was introduced for the first time in
\cite{Rosenbaum:1998}, being called $g^{(2)}$. The
generators $L^i,i=1,\ldots,7$, generate a subalgebra of the form  $gl_2 \ltimes R^3$. For each $n\in\mathbb{N}$,  the generators (\ref{g2_gen}) have the
common invariant subspace
\begin{equation}
\label{g2_inv}
 {P}^{(2)}_{n, \{1,2 \}}\ =\ \langle\tau_1^{n_1}\tau_2^{n_2} |\ 0\leq (n_1+2 n_2)
\leq n\rangle \ ,
\end{equation}
(cf. Definition 2), on which they act irreducibly \footnote{It is also  worth
mentioning that at $n=0$, the algebra $gl_2\ltimes R^3$ becomes an algebra of vector fields, acting on a 2-Hirzebruch surface, $\Si_2$, and the modules are the sections of
holomorphic line bundles over this surface (see \cite{ghko} and
references therein).}. The common invariant spaces ${P}^{(2)}_{n,
\{1,2 \}}, n\in \mathbb{N}$ form the infinite flag ${P}^{(2)}_{\{1,2 \}}$. If
the generator $L^4$ is excluded, the remaining generators preserve ${P}^{(2)}_{\{1,2 \}}$.

It can be easily shown that the operator (\ref{h_tau_G2}) can be
rewritten in terms of the generators of the algebra $g^{(2)}$, in the
$n=0$ case, as
\[
h_{G_2} (\tau)\ =\ 4L^1 L^1 + L^2 L^1 + \frac{1}{3} T -
\frac{1}{3}L^2 L^2 - 12 L^1 L^5 -2 L^1 L^3
\]
\[
-\frac{1}{2} L^2 L^3 +2 L^6 L^2-9 L^5 L^6 - \frac{3}{2} L^3 L^5 -
\frac{3}{2} L^3 L^6 - \frac{1}{4} L^3 L^3 + L^6 L^7
\]
\begin{equation}
+ 2\nu L^1 - \frac{3\mu + 4 \nu}{3}L^2 - 3(2\mu+\nu)L^5 -
\frac{1+4\mu+4\nu}{4}L^3 - \frac{\nu}{12}L^7 \ .
\end{equation}
In this representation, the generator $L^4$ is absent. Hence, the algebra $g^{(2)}$ is the hidden algebra of the $G_2$ trigonometric model.

\section{The case $\De = F_4$}

The trigonometric $F_4$ model is defined by the
Olshanetsky-Perelomov Hamiltonian (\ref{H}) for the root space
$F_4$
\footnote{Actually, this form corresponds to the
representation of the $F_4$-Hamiltonian for the dual root space (see
a discussion in \cite{blt}). It was chosen for convenience in making calculations. Calculations with the $F_4$-Hamiltonian defined for the  root space turned to be more complicated.},
\[
 {H}_{\rm F_4}  = -\frac{1}{2} \sum_{i=1}^{4}
\partial_{x_i}^2 + \frac{g\beta^2}{4}\ \sum_{j>i}
 \left(
\frac{1}{\sin^2 \frac{\beta(x_i-x_j)}{2}} +
\frac{1}{\sin^2 \frac{\beta(x_i+x_j)}{2}}
 \right)
\]
\begin{equation}
\label{H_F4}
 + g_1 \beta^2\ \sum_{i=1}^{4}\frac{1}{\sin^2 \beta {x_i}} +
g_1\beta^2\ \sum_{ \nu's=0,1} \frac{1}{\sin^2 \frac{\beta \left[ x_1
+ (-1)^{\nu_2}x_2+ (-1)^{\nu_3}x_3+ (-1)^{\nu_4}x_4 \right]}{2}} \ .
\end{equation}
where $\beta$ is a parameter and $g, g_1 > -1/4$ are the coupling
constants. If $g_1=0$ the Hamiltonian (\ref{H_F4}) degenerates to
that of the trigonometric $D_4$ model (i.e. (\ref{H_BCN}) at $N=4$ and
$g_2=g_3=0$). The trigonometric $F_4$ model is completely integrable, for arbitrary values of the coupling constants $g,g_1$. It describes a quantum particle in a
four-dimensional space.

The ground state of the  Hamiltonian (\ref{H_F4}) is
\begin{equation}
\label{Psi_F4}
 \Psi_0= \Delta_-^{\nu}(\beta)\Delta_+^{\nu}(\beta)\Delta_0^{\mu}(\beta)
 \Delta^{\mu}(\beta) \ ,
\end{equation}
where degrees $\nu, \mu$ are related with the coupling constants by
\begin{equation}
g = \nu(\nu-1)\ , \qquad g_1 = \frac{1}{2}\mu(\mu-1)\ ,
\end{equation}
and
\begin{subequations}
\begin{eqnarray}
\De_{\pm}(\beta) &=& \prod^4_{j<i}\sin
\frac{\beta(x_i\pm x_j)}{2}\ , \\  \De_{0}(\beta) &=&
 \prod^4_{i=1}\sin \beta x_i\ ,  \\ \De (\beta) &=&
\prod_{\nu's}\sin \frac{\beta \left[x_1 +
(-1)^{-\nu_2}x_2 + (-1)^{-\nu_3}x_3+ (-1)^{-\nu_4}x_4 \right]}{2}\ .
\end{eqnarray}
\end{subequations}
The ground state eigenvalue is
\begin{equation}
\label{E0_F4}
 E_0\ =\ (7 \nu^2 + 14 \mu^2 + 18
\nu\mu)\beta^2 \ .
\end{equation}

In \cite{blt}, as a result of a series of intelligent guesses, after rather
sophisticated and tedious analysis there were  found
surprisingly simple variables in terms of which the similarity-transformed
version of the Hamiltonian (\ref{H_F4}),
\begin{equation}
\label{h_F4}
 h_{\rm F_4} \ =\ -2(\Psi_{0})^{-1}({
 H}_{\rm F_4}-E_0)(\Psi_{0}) \ ,
\end{equation}
takes the form of an algebraic operator. This operator was derived
explicitly (see \cite{blt}). It preserved the infinite flag ${P}^{(4)}_{\{ 1,2,2,3 \}}$.
This is in agreement with our
conjecture that the characteristic vector for a trigonometric model always coincides
with the minimal characteristic vector for the corresponding rational model~\cite{BLT-rational}. The explicit
expressions for the abovementioned  variables are\footnote{In the limit $\beta$
tends to zero the variables $\eta$'s go to the polynomial invariants
of the $F_4$ root space which are classified following the degrees
of the $F_4$ algebra $(2,6,8,12)$. Numbering of $\eta$'s variables
reflects this fact.}
\begin{subequations}
\begin{eqnarray}
\label{eta_F4}
 {{\eta}_{2}} &=& {{\widetilde\eta}}_{1}-\frac{\beta^2}{6}{{\widetilde\eta}}_{2}\ ,  \\
 {{\eta}_{6}} &=&  {{\widetilde\eta}_3} - {\frac
{1}{6}} \, {{\widetilde\eta}_1}\,{{\widetilde\eta}_2}-
\frac{\beta^2}{2}({{\widetilde\eta}}_{4}-\frac{1}{36}
{{\widetilde\eta}}_{2}^2)  , \\
 {{ \eta}_{8}} &=& {{\widetilde\eta}_4} - {
\frac{1}{4}} \, {{\widetilde\eta}_1}\,{{\widetilde\eta}_3} +
{\frac{1}{12}} \, {{\widetilde\eta}_2}^{2}  ,\\
 {{ \eta}_{12}} &=&
{{\widetilde\eta}_4}\,{{\widetilde\eta}_2} - {\frac{1 }{36}}
\,{{\widetilde\eta}_2}^{3} - {\frac {3}{8}} \,
{{\widetilde\eta}_3}^{2} + {\frac{1}{8}} \,{{\widetilde\eta}_1}\,
{{\widetilde\eta}_2}\,{{\widetilde\eta}_3} - {\frac{3}{8}} \,
{{\widetilde\eta}_1}^{2}\,{{\widetilde\eta}_4}\ ,
\end{eqnarray}
\end{subequations}
where $\tilde \eta$'s are the elementary symmetric polynomials
\begin{equation}
 {\widetilde\eta}_i\ =\ \si_i\left(\frac{4\sin^2
\frac{\beta}{2} x}{\beta^2} \right)\ ,\ i=1,2,3,4\ ,
\end{equation}
(For a definition of $\si$'s, see (\ref{RT_A})). Below we shall show
that there is nothing mysterious in the variables (\ref{eta_F4});
they can easily be obtained from the FTI of (\ref{Trig_Inv}).

Associated with the algebra $F_4$ are a root space and
a dual-root space. From a technical point of view, it is more convenient
to work in the dual root space. Thus, we shall define the FTI by averaging over  orbits in the dual root space. In the dual root space, the fundamental weights
\begin{equation}
\label{F4_FW}
 a_1= e_3+e_4\ ,\ a_2= 2 e_4\ ,\ a_3= e_2+e_3+2e_4\ ,\ a_4=
 e_1+e_2+e_3+3e_4\ ,
\end{equation}
generate orbits with lengths equal to $(24,24,96,96)$, respectively.
Averaging over the orbits, as in (\ref{Trig_Inv}), we define FTI that we denote
by $\tau_{1,2,3,4}$, respectively.
After some algebra one finds an explicit relation between $\eta$'s
and $\tau$'s:
\begin{subequations}
\label{transf_F4}
\begin{eqnarray}
 && \quad \eta_2      =  -\frac{1}{24}\frac{\tau_1-24}{\beta^2}, \\
 && \quad \eta_6      =   \frac{1}{4608}\frac{\tau_1^2+24\tau_1-36\tau_2-288}{\beta^6}, \\
 && \quad \eta_8      =   \frac{1}{3072}\frac{\tau_1^2-12\tau_1-3\tau_3}{\beta^8},\\
 && \quad \eta_{12} =   -\frac{1}{294912}\frac{2\tau_1^3+72\tau_1^2-9\tau_1 \tau_3-864\tau_1-
  324\tau_2-216\tau_3+27\tau_4-1728}{\beta^{12}}.
\end{eqnarray}
\end{subequations}
It is evident that this transformation leads to an algebraic form for $h_{\rm F_4}$; and in fact, this algebraic form  preserves the infinite flag ${P}^{(4)}_{\{
1,2,2,3 \}}$. The form of this operator in terms of the trigonometric
invariants (the $\tau$'s)  is the following
\begin{equation}
\label{h_F4_tau}
 {h}_{F_4}(\tau)\ \equiv \ \frac{1}{4\beta^2}{h}_{F_4}\ =\
 \sum_{i,j=1}^{4} {A}_{ij}({\tau})
 \frac{\pa^2}{\pa {{\tau}_i} \pa {{\tau}_j} } +
 \sum_{i=1}^{4}{B}_i({\tau}) \frac{\pa}{\pa {\tau}_i} \ ,
\end{equation}
where
\[
A_{ 1 1 }=  - 2\tau_1^2 + 24\tau_1 + 12\tau_2+ 2\tau_3 + 96\ , \
A_{ 1 2 }= - 2\tau_1 \tau_2 + 24\tau_1+ 6\tau_3\ ,
\]
\[
A_{ 1 3 }=  24\tau_1^2 + 8 \tau_1\tau_2 - 3\tau_1\tau_3
    - 192\tau_1- 84\tau_2 - 48\tau_3 + 3\tau_4 - 576\ ,
\]
\[
A_{ 1 4 }=  8\tau_1\tau_2 - 4\tau_1\tau_4  + 4\tau_2\tau_3 - 96\tau_1
    - 24\tau_3 \ ,
\]
\[
A_{ 2 2 }= 24\tau_1^2 - 4\tau_2^2  - 192\tau_1 - 96\tau_2 - 48\tau_3
     + 4\tau_4 - 384\ ,
\]
\begin{eqnarray*}
A_{ 2 3 } &=& -48\tau_1^2- 8\tau_1\tau_2 + 6\tau_1 \tau_3 - 4\tau_2\tau_3
  + 480\tau_1 + 216\tau_2+ 120\tau_3 - 18\tau_4 + 1152\ ,
\end{eqnarray*}
\begin{eqnarray*}
A_{ 2 4 } &=&  - 48\tau_1^3 - 8\tau_1^2\tau_2 \\&&
   + 192\tau_1^2 +208\tau_1\tau_2 + 144\tau_1\tau_3- 12\tau_1 \tau_4
   + 24\tau_2^2+16\tau_2\tau_3 - 6\tau_2 \tau_4+ 6\tau_3^2  \\&&
   + 3072\tau_1+960\tau_2+576\tau_3- 96\tau_4+ 4608\ ,
\end{eqnarray*}
\begin{eqnarray*}
A_{ 3 3 }&=& 24\tau_1^3+ 8\tau_1^2\tau_2- 192\tau_1^2 - 120\tau_1\tau_2
    - 72\tau_1\tau_3+ 2\tau_1\tau_4 -8\tau_2\tau_3 - 6\tau_3^2\\ &&
    - 768\tau_1 - 96\tau_2 - 96\tau_3 + 24\tau_4 \ ,
\end{eqnarray*}
\begin{eqnarray*}
A_{ 3 4 } &=& 4\tau_1\tau_2\tau_3-32\tau_1^2\tau_2 + 192\tau_1^2
   + 288\tau_1\tau_2 - 24\tau_1\tau_3-16\tau_1\tau_4 \\ &&
   + 144\tau_2^2 + 64\tau_2\tau_3-12\tau_2\tau_4 - 8\tau_3\tau_4
   - 1920\tau_1-96\tau_2-480\tau_3+72\tau_4- 4608\ ,
\end{eqnarray*}
\begin{eqnarray*}
A_{ 4 4 } &=& - 32\tau_1^3\tau_2 - 384\tau_1^3 - 192\tau_1^2 \tau_2
   -16\tau_1^2\tau_4 + 96\tau_1\tau_2^2 + 4\tau_2\tau_3^2 \\  && +96\tau_1\tau_2\tau_3-8\tau_1\tau_2\tau_4  + 2688\tau_1^2
   + 1728\tau_2^2 + 48\tau_3^2 - 12\tau_4^2 \\ &&
   + 5760\tau_1\tau_2 + 1152\tau_1\tau_3 + 32\tau_1\tau_4
   + 1024\tau_2\tau_3 - 48\tau_2\tau_4 + 32\tau_3\tau_4 \\ &&
   + 15360\tau_1 + 12288\tau_2 + 2304\tau_3 + 192\tau_4 + 18432\ ,
\end{eqnarray*}
and
\[
B_1 =  - 2(1+6\,\mu + 5\,\nu)\tau_1 - 48\,\nu\ ,
\]
\[
B_2 = - 12\,\nu \,\tau_1 - 4(1+5\,\mu + 3\,\nu)\,\tau_2 -
96\,\mu\ ,
\]
\[
B_3 = - 48(\mu \, + \nu )\,\tau_1 - 24\,\nu \,\tau_2 - 6
(1+4\,\mu + 3\,\nu)\,\tau_3 \ ,
\]
\begin{eqnarray*}
B_4 &=&  - 48\,\mu \,\tau_1^2 - 8\,\nu \,\tau_1\tau_2 +
48\,(8\,\mu + \,\nu)\,\tau_1 + 48\,(4\,\mu - \,\nu)\,\tau_2 \\
&& + 96\,\mu \,\tau_3 - 12\,(1+3\,\mu + 2\,\nu)\,\tau_4 +
1152\,\mu \ .
\end{eqnarray*}

A straightforward analysis confirms the conclusion that the operator ${h}_{F_4}
(\tau)$ preserves the infinite flag ${P}^{(4)}_{\{ 1,2,2,3 \}}$. This is not a surprising result, since the transformation (\ref{transf_F4}) maps each subspace
in ${P}^{(4)}_{\{ 1,2,2,3 \}}$ to itself.

The $F_4$ model depends on two parameters $\nu, \mu$, and
the nodal structure of the eigenpolynomials (i.e. where they vanish) at fixed
$\nu$ and $\mu$ remains an open question.

This Hamiltonian can be written in terms of the generators of an infinitedimensional algebra of differential operators  $f^{(4)}$
generated by 49 operators, which admits finite-dimensional
representations in terms of inhomogeneous polynomials in four
variables (see \cite{blt}). Among those 49 operators there are 22
differential operators of the first order, 22 of the second
and 5 the of third.

\section{The case $\De = E_6$}

The Hamiltonian of the trigonometric $E_6$ model is built using the
root system of the $E_6$ algebra (see (\ref{H})). A convenient way
to represent the Hamiltonian in  coordinate form is to use an
$8$-dimensional space with coordinates $x_1,x_2,\ldots x_8$ imposing
two constraints: $x_7=x_6, \, x_8=-x_6$. In terms of these coordinates,
\begin{equation}
\label{H_E6}
  {H}_{E_6} = -\frac{1}{2} \Delta^{(8)} + \frac{g \beta^2}{4} \sum_{j<i =1}^{5}
  \left[\frac{1}{\sin^{2} {\frac{\beta}{2}(x_i + x_j)}} + \frac{1}{\sin^{2}
  {\frac{\beta}{2} (x_i - x_j)}} \right]
\end{equation}
\[
+ \frac{g \beta^2}{4} \sum_{\{\nu_j\}} \frac{1}{\left[\sin^{2}
\dfrac{\beta}{4}\left({ -x_8 + x_7 +x_6 - \sum_{j=1}^5
(-1)^{\nu_j}x_j}\right)\right]} ~, \
\]
the second summation being one over quintuples $\{\nu_j\}$ where each $\nu_j = 0, 1,$ and   $\sum_{j=1}^{5} \nu_j \text{~is even}$. Here
$g=\nu(\nu-1)>-1/4$ is the coupling constant. The configuration space is the principal $E_6$ Weyl alcove.

In order to resolve the constraints, we introduce new variables:
\begin{eqnarray}
\label{E6_y-vars}
 y_i &=& x_i\ , \quad i=1\ldots 5 \non \\
 y_6 &=& x_6 + x_7 - x_8\ , \qquad  \mbox{
(with the constraint $y_6=3x_6$)}, \non\\
 y_7 &=& x_6 -x_7\ ,\qquad \mbox{(with the constraint $y_7=0$)},
\non\\
 y_8 &=& x_6 +x_8\ ,\qquad \mbox{(with the constraint
$y_8=0$)}.
\end{eqnarray}
In terms of these, the Laplacian has the representation
\begin{equation}
\label{E6_Laplace}
 \Delta^{(8)} = \Delta_y^{(5)} + 3 \frac{\pa^2}{\pa y_6^2} +
 2\left[
  \frac{\pa^2}{\pa y_7^2} + \frac{\pa^2}{\pa y_8^2} +
  \frac{\pa^2}{\pa y_7 \pa y_8}
  \right] \ ,
\end{equation}
while the potential part of (\ref{H_E6}) depends on $y_1 \ldots y_6 $ only:
\begin{eqnarray}
\label{E6_potential}
 V &=&   \frac{g \beta^2}{4} \, \sum_{j<i =1}^{5}
 \left[
\frac{1}{\sin^{2} \frac{\beta}{2}(y_i + y_j)^2} + \frac{1}{\sin^{2}
 \frac{\beta}{2}(y_i - y_j)} \right] \non \\
&& \, + \, \frac{g\beta^2}{4} \sum_{\nu_j,j=1}^5
 \frac{1}{\left[\sin^{2} \frac{\beta}{4}\left({ y_6 - \sum_{j=1}^{5}
(-1)^{\nu_j}y_j }\right)\right]}\ .
\end{eqnarray}
In this formalism, imposing the constraints
requires that one should study only eigenfunctions having no dependence on
$y_7, y_8$. Hence, the $y_{7,8}$-dependent part of the Laplacian standing in square brackets in (\ref{E6_Laplace}) can  simply be dropped.

The ground state eigenfunction and its eigenvalue are
\begin{equation}
\label{Psi_E6}
 \Psi_0 = (\De_+^{(5)} \De_-^{(5)})^\nu \De_{E_6}^\nu
 \ ,
 \ E_0 = 39 \beta^2 \nu^2\ ,
\end{equation}
where
\begin{subequations}
\begin{eqnarray}
 \De_\pm^{(5)} &=& \prod_{j<i =1}^{5} \sin \frac{\beta}{2} (y_i \pm y_j)\ ,\\
 \De_{E_6} &=& \prod_{\{\nu_j\}}
 \sin \frac{\beta}{4}(y_6 + \sum_{j=1}^{5} (-1)^{\nu_j}y_j)\ .
\end{eqnarray}
\end{subequations}
The main object of our study is the similarity-transformed
version of the Hamiltonian (\ref{H_E6}), with the ground state eigenfunction
(\ref{Psi_E6}) taken as a factor, i.e.
\begin{equation}
\label{h_E6}
 h_{\rm E_6} \ =\ -\frac{8}{\beta^2}(\Psi_{0})^{-1}({
 H}_{\rm E_6}-E_0)(\Psi_{0}) \ ,
\end{equation}
where $E_0$ is given by (\ref{Psi_E6}).

The $E_6$ root space is characterized by 6 fundamental weights, which
generate orbits of  lengths ranging from 27  to 720. Let us
introduce an ordering of the fundamental trigonometric invariants $\tau_a$ defined by
(\ref{Trig_Inv}), namely:
\[
\begin{array}[h]{lcl}
\mbox{orbit Variable}    &  \mbox{weight vector} &  \mbox{orbit
size}
\\
\tau_1    &             -2e_6          &27
\\
\tau_2    &             e_5-e_6         &27
\\
\tau_3    &        e_4+e_5-2e_6          &216
\\
\tau_4    &   -\frac{1}{2}(e_1-e_2-e_3-e_4-e_5)  -\frac{5}{2} e_6
&216
\\
\tau_5    &    \frac{1}{2}(e_1+e_2+e_3+e_4+e_5)  -\frac{3}{2} e_6
& 72
\\
\tau_6    &                  e_3+e_4+e_5-3e_6          & 720
\end{array}
\]
The pairs of variables $(\tau_1,\tau_2)$ and $(\tau_3,\tau_4)$ are complex
conjugates. The orbit variables have  certain transformation properties under the involution $\beta \rar - \beta$: $\tau_{1,3}(-\beta) = \tau_{2,4}(\beta)$, while
$\tau_{5,6}$ remain unchanged, i.e. are invariant. Since the Hamiltonian is invariant under $\beta \rar - \beta$, after converting to  $\tau$-variables
it should be invariant under the simultaneous interchange $\tau_{1} \lrar \tau_{2}, \tau_{3} \lrar \tau_{4}$.

After very lengthy, truly cumbersome and tedious calculations which
probably will be published elsewhere, one can show that
the similarity-transformed Hamiltonian (\ref{h_E6}), in terms of the above
trigonometric invariants ($\tau$-variables), takes on an algebraic form. 
This is the following:
\begin{equation}
\label{h_E6_tau}
 {h}_{E_6}\ =\
 \sum_{i,j=1}^{6} {A}_{ij}({\tau})
 \frac{\pa^2}{\pa {{\tau}_i} \pa {{\tau}_j} } +
 \sum_{i=1}^{6}{B}_i({\tau}) \frac{\pa}{\pa {\tau}_i} \ ,
\end{equation}
where
\[
A_{1 1}= -\frac {4 \tau_{1}^{2}}{3}  + 20 \tau_{2}\, + 2 \tau_{4}
\,,\quad
A_{1 2} = -\frac {2 \tau_{1} \tau_{2}}{3} + 6 \tau_{5} + 54 \,,
\]
\[
A_{1 3} = -\frac {4 \tau_{1} \tau_{3}}{3} + 5
\tau_{2}\tau_{5}  - 32 \tau_{2}- 5 \tau_{4}
\,,\quad
A_{1 4} =  16 \tau_{1} \tau_{2} - {  \frac {5 \tau_{1}
\tau_{4}}{3}}  - 51 \tau_{5} + 3 \tau_{6}  - 432 \,,
\]
\[
A_{1 5}= - \tau_{1} \tau_{5} + 32 \tau_{1}  + 5 \tau_{3}
\,,\quad
A_{1 6}= 10\tau_1\tau_5 - 2 \tau_1 \tau_6 - 64 \tau_2^2- 4
\tau_2\tau_4 + 4\tau_{3} \tau_{5} + 384 \tau_1+ 78 \tau_3 \,,
\]
\[
A_{2 2}=-\frac {4 \tau_{2}^{2}}{3} + 20 \tau_{1}  + 2 \tau_{3}
\,,\quad
A_{2 3}= 16 \tau_{1}\tau_{2} - { \frac{5\tau_{2}\tau_{3}}{3}}
- 51 \tau_{5}  + 3 \tau_{6} - 432 \,,
\]
\[
A_{2 4}=-\frac {4 \tau_{2} \tau_{4}}{3} + 5 \tau_{1} \tau_{5}
- 32 \tau_{1} - 5 \tau_{3}
\,,\quad
A_{2 5}= - \tau_{2} \tau_{5} + 32 \tau_{2}  + 5 \tau_{4} \,,
\]
\[
A_{2 6}= - 64 \tau_{1}^{2} - 4 \tau_{1} \tau_{3} + 10 \tau_{2} \tau_{5} - 2 \tau_{2} \tau_{6}
+ 4 \tau_{4} \tau_{5} + 384 \tau_{2} + 78 \tau_{4} \,,
\]
\[
A_{3 3} =  16 \tau_{1} \tau_{2}^{2} - 64 \tau_{1}^{2}
- 24 \tau_{1} \tau_{3} - 36 \tau_{2} \tau_{5} + 2 \tau_{2} \tau_{6}
- \frac {10 \tau_3^2}{3}+ 4 \tau_{4} \tau_{5}
  - 208 \tau_{2}  + 8 \tau_{4} \,,
\]
\[
A_{3 4}=  4 \tau_1\tau_2\tau_5 - 176 \tau_1\tau_{2}
- \frac{8\tau_3\tau_4}{3} + 6\tau_5^2 + 528\tau_5- 42\tau_6+ 3888\,,
\]
\[
A_{3 5}= 32 \tau_{2}^{2}  + 4 \tau_{2} \tau_{4}  - 2 \tau_{3} \tau_{5}
 - 224 \tau_{1} - 44 \tau_{3}\,,
\]
\begin{eqnarray*}
A_{3 6} &=& - 128 \tau_1^2\tau_2 + 5\tau_1\tau_5^2 + 3\tau_2\tau_4\tau_5
    + 320 \tau_1\tau_5 - 32 \tau_1\tau_6  \\ &&
  + 576 \tau_2^2 + 104 \tau_2\tau_4
  - \tau_3\tau_5 - 4\tau_3\tau_6+5\tau_4^2+192\tau_1-312\tau_3 \,,
\end{eqnarray*}
\[
A_{4 4}=  16 \tau_{1}^{2} \tau_{2} - 36 \tau_{1} \tau_{5}    + 2 \tau_{1} \tau_{6} - 64 \tau_{2}^{2} - 24 \tau_{2} \tau_{4}
 + 4 \tau_{3} \tau_{5} - \frac {10\tau_4^2}{3} - 208 \tau_{1} +8 \tau_{3}\,,
\]
\[
A_{4 5}=  32 \tau_{1}^{2} + 4 \tau_{1} \tau_{3}
- 2 \tau_{4} \tau_{5}  - 224 \tau_{2} - 44 \tau_{4} \,,
\]
\begin{eqnarray*}
A_{4 6} &=& -128\tau_1\tau_2^2  + 3 \tau_1\tau_3\tau_5 + 5\tau_2\tau_5^2
     + 576\tau_1^2 + 104\tau_{1}\tau_{3} +320
\tau_{2}\tau_{5}
\\ &&
 - 32\tau_2\tau_6+5\tau_3^2-\tau_4\tau_5-4\tau_4\tau_6 + 192 \tau_{2}
- 312 \tau_{4} \,,
\end{eqnarray*}
\[
A_{5 5}= 16\tau_1\tau_2- 2\tau_5^2- 36\tau_{5}  + 2\tau_{6}  - 144\,,
\]
\[
A_{5 6}= - 96\tau_1\tau_2 + 3 \tau_3\tau_4+ 15 \tau_5^2 - 3\tau_5\tau_6
       +216\tau_5-12\tau_6 + 864\,,
\]
\begin{eqnarray*}
A_{6 6} &=& - 64\tau_1^2\tau_2^2 + 4\tau_1\tau_{2}\tau_{5}^{2}
 + 256\tau_{1}^3+ 32\tau_1^2\tau_3+ 80\tau_1\tau_2\tau_5
 - 24 \tau_1\tau_2\tau_6 + 4\tau_1\tau_3^2
 \\ && \hspace{-30pt}
  - 16\tau_1\tau_4\tau_5 + 256\tau_2^3 + 32 \tau_2^2\tau_4
  - 16 \tau_2\tau_3\tau_5+4\tau_2\tau_4^2 + 2\tau_3\tau_4\tau_5
  + 6\tau_5^3-2112\tau_1\tau_2
  \\ && \hspace{-30pt}
  - 96\tau_1\tau_4- 96\tau_2\tau_3+ 84\tau_3\tau_4+216\tau_5^2
  + 36\tau_5\tau_6 - 6 \tau_6^2
  + 2592\tau_5 + 288\tau_6 + 10368 \,,
\end{eqnarray*}

\noindent
 and

\[
B_{1} =  - \frac {4(6+\nu)}{3}\tau_1 \ ,\ B_2 =  - \frac {4(6+\nu)
}{3} \tau_2 \ ,
\]
\[
B_3\, = \, -\frac{1}{18} [(1-\nu) ({\tau_1}^2 + 5 \tau_1\,\tau_2
 +9{\tau_2}^2 - 15 \tau_2  - 54 \tau_4 - 45 \tau_5 - 405)
\]
\[
  +30 (13+3\nu) \tau_1 + (171+ 49\nu) \tau_3]\ ,
\]
\[
B_4\, = \, -\frac{1}{18}[(1-\nu)(9{\tau_1}^2 + 5 \tau_1\,\tau_2 + {\tau_2}^2
  - 15 \tau_1  - 54 \tau_3 - 45 \tau_5 - 405)
\]
\[
 \ +30 (13+3\nu) \tau_2 + (171+ 49\nu) \tau_4]\ ,
\]
\[
{B}_{5}\, = -\frac{1}{108}(1 - \nu)[2 ({\tau_1}^2 + \tau_1\,\tau_2 + {\tau_2}^2) - 30 (\tau_1 + \tau_2) - 3(\tau_3 + \tau_4)]
\]
\[
 -\frac{13 (5 + \nu)}{6} \tau_5 - \frac{3}{2}(47 + \nu) \ ,
\]
\[
 {B}_{6}\, =  \frac{(1-\nu)}{108}  [
  2({\tau_1}^3 + 2 {\tau_1}^2\tau_2+ 2\tau_1 {\tau_2}^{2} + {\tau_2}^3) -3(26 {\tau_1}^2 +9\tau_1\tau_3 + 11\tau_1\tau_4 +12\tau_1\tau_5
\]
\[
  +26{\tau_2}^2 +11\tau_2\,\tau_3 +9\tau_2\tau_4+12\tau_2\tau_5)+ 3438 (\tau_1 + \tau_2) + 522 (\tau_3 + \tau_4)]
\]
\[
 - \frac{(43 +29\nu)}{3} \tau_1\tau_2 + (23+61\nu) \tau_5 - (20 +7\nu) \tau_6 +108(1+5\nu) \ .
\]
After some analysis, one finds that the operator (\ref{h_E6_tau}) preserves
the infinite flag ${P}^{(6)}_{\{ 1,1,2,2,2,3 \}}$. Its  characteristic vector
$\vec \alpha = (1,1,2,2,2,3)$  coincides with the minimal
characteristic vector for the corresponding rational model
\cite{BLT-rational}. This confirms our conjecture that the characteristic
vector for a trigonometric model always coincides
with the minimal characteristic vector for the corresponding rational model.

It is worth mentioning that the operator (\ref{h_E6_tau}) has a symmetry with respect to FTI (orbit variables) generated by  orbits of the same length (see above); i.e.,
$\tau_1 \lrar \tau_2, \tau_3 \lrar \tau_4$. Under this involution
\[
  A_{12} \lrar A_{12}\ ,\
  A_{13} \lrar A_{24}\ ,\ A_{14} \lrar A_{23}\ ,\ A_{15} \lrar A_{25}\ ,\ A_{16} \lrar A_{26}\ ,\ A_{33} \lrar A_{44}\ ,\
\]
\[
  A_{34} \lrar A_{34}\ ,\
  A_{35} \lrar A_{45}\ ,\ A_{36} \lrar A_{46}\ ,\
  A_{55} \lrar A_{55}\ ,\ A_{56} \lrar A_{56}\ ,\ A_{66} \lrar A_{66}\ ,\
\]
\[
  B_1 \lrar B_2\ ,\ B_3 \lrar B_4\ ,\ B_5 \lrar B_5\ ,\ B_6 \lrar B_6\ .
\]

The $E_6$ model depends on the parameter $\nu$, and
the nodal structure of eigenpolynomials (i.e. where they vanish) at fixed
$\nu$ remains an open question.

\section{Summary and conclusions}

Weyl-invariant coordinates leading to the algebraic forms of the 
trigonometric Olshanetsky-Perelomov Hamiltonians associated to the crystallographic root spaces $A_N, BC_N, G_2, F_4$ were found (in  \cite{Ruhl:1995}, \cite{Brink:1997}, \cite{Rosenbaum:1998}, \cite{blt}, respectively) in a manner that was specific to each problem. In this paper, we have shown that the fundamental trigonometric invariants (FTI), if used as coordinates, provide a systematic way of reducing the trigonometric Hamiltonians associated to $A_N, B_N, C_N, D_N, BC_N$,
and $G_2, F_4, E_6$, to algebraic form. The eigenfunctions of the
trigonometric Hamiltonians (i.e., the Jack polynomials) remain polynomials 
in the FTI. The use of FTI enabled us to find an algebraic form of the Hamiltonian associated to $E_6$, which did not seem feasible at all, in the past. The calculations in this paper were based on a straightforward change of variables from Cartesian coordinates to FTI. Actually, there are clear indications of the existence of a representation-theoretic
formalism that may allow such results to be derived more rapidly and elegantly \cite{Ruehl:1999,Sasaki:2000,Nekrasov}.

Each of the Olshanetsky-Perelomov Hamiltonians, in algebraic form, 
preserves an infinite flag of polynomial spaces, with a characteristic vector ${\vec \alpha}$ that coincides with the minimal characteristic vector for the corresponding rational model (cf. \cite{BLT-rational}).
It is worth noting that the matrices $A_{ij}$ in the algebraic form Hamiltonians given explicitly in Eqs. (2.6), (3.4), (4.8), (5.11), (6.8), with polynomial entries, correspond to flat-space metrics, in the sense that the associated Riemann tensor vanishes. The change of variables in the corresponding Laplace-Beltrami operator, from FTI to
Cartesian coordinates, transforms these metrics to diagonal form.

It should be stressed that each Hamiltonian of the form (\ref{H})  is completely integrable. This implies the existence of a number of operators (the `higher Hamiltonians') which commute with it and which are in involution. It is evident that these commuting operators take on an algebraic form after a gauge rotation (with the corresponding ground state eigenfunction as a gauge factor), and a change of variables from Cartesian coordinates to the FTI, i.e., to the $\tau$'s. Although both the original Hamiltonian (\ref{H})  and the FTI (\ref{Trig_Inv}) depend on the real parameter $\beta$, the resulting algebraic forms are $\beta$-independent. This fact yields a non-trivial connection between the algebraic operators of trigonometric models and the corresponding rational models. In practice,
the connection is made in the following way: (i) take the set of FTI, specially ordered; and (ii) subtract from each a certain nonlinear combination of the other FTI, in such a way that as $\beta \to 0$, one obtains the polynomial Weyl invariants, which in the rational case lead to an algebraic operator, preserving a minimal flag. Surprisingly, if one changes variables to these transformed FTI, the operator, expressed in terms of them, remains in algebraic form. This makes it possible to derive the algebraic operator of the
rational model by taking the $\beta \to 0$ limit \cite{BLT-rational}. The significance of the ordering of the FTI, on which this procedure depends, is not clear to the authors.
An analysis similar to the analysis of this paper has not yet been presented for the case of the trigonometric Olshanetsky-Perelomov Hamiltonians related to the exceptional root spaces $E_7$ and $E_8$. We conjecture that in these cases as well, the FTI taken as coordinates will yield an algebraic form for the Hamiltonian, and that the infinite flag of polynomial spaces with the same characteristic vector as in the corresponding rational model will be preserved. In concluding, we mention that the existence of algebraic forms of Olshanetsky- Perelomov Hamiltonians makes possible the study of their perturbations by purely algebraic means: one can develop a perturbation theory in which all corrections are found by linear-algebraic methods \cite{Tur-pert}.
It also gives a hint that quasi-exactly-solvable generalizations of the Olshanetsky-Perelomov Hamiltonians may exist.

\bigskip

\textit{\small Acknowledgements}. The computations in this paper were performed on MAPLE 8 with the package COXETER created by J.~Stembridge. One of us (A.V.T.) is grateful to IHES for its kind hospitality extended to him, while the paper was completed. A.V.T. also thanks Prof. N. Nekrasov for valuable discussions.

\bibliographystyle{amsalpha}

\end{document}